JOURNAL OF COMPUTING, VOLUME 2, ISSUE 2, FEBRUARY 2010, ISSN 2151-9617    142
HTTPS://SITES.GOOGLE.COM/SITE/JOURNALOFCOMPUTING/# Optimization Digital Image Watermarking Technique for Patent Protection

Mahmoud Elnajjar, A.A Zaidan, B.B Zaidan, Mohamed Elhadi M.Sharif and Hamdan.O.Alanazi***Abstract***— The rapid development of multimedia and internet allows for wide distribution of digital media data. It becomes much easier to edit, modify and duplicate digital information besides that, digital documents are also easy to copy and distribute, therefore it will be faced by many threats. It is a big security and privacy issue. Another problem with digital document and video is that undetectable modifications can be made with very simple and widely available equipment, which put the digital material for evidential purposes under question With the large flood of information and the development of the digital format, it become necessary to find appropriate protection because of the significance, accuracy and sensitivity of the information ,therefore  multimedia technology and popularity of internet communications they have great interest in using digital watermarks for the purpose of copy protection and content authentication. Digital watermarking is a technique used to embed a known piece of digital data within another piece of digital data .A digital data may represent a digital signature or digital watermark that is embedded in the host media. The signature or watermark is hidden such that it's perceptually and statistically undetectable. Then this signature or watermark can be extracted from the host media and used to identify the owner of the media.

## 1. INTRODUCTION

Digital document can be distributed via the World Wide Web to a large number of people in a cost–effective way, but the increasing importance of digital media brings a new challenges as it now straightforward duplicate and even manipulate multimedia content this give a strong need for security services in order to keep the distribution of digital multimedia work both profitable for the document owner and reliable for the customer [1],[2]. Digital data can easily be exactly copied; this very useful but it also poses problems such as detect their values, for example, digital images, or record music digital, replacing a given piece of digital data cannot be distinguished and their pedigree cannot be confirmed[3],[4],[5]. It's impossible to determine which piece is the original and which is the copy [6]. Techniques that use a copying of data are generally easy to defeat once a clever individual discover the secret algorithm. The legality of defeating such techniques is debatable but the weakness of the technical approach is not [7],[8].

————————————

- *Mahmoud Elnajjar PhD- Candidate /Universiti Utara Malaysia / Microsoft Business Intelligence MCSD.NET  Developer Solution.*
- *A. A. Zaidan – PhD Candidate on the Department of Electrical & Computer Engineering , Faculty of Engineering , Multimedia University , Cyberjaya, Malaysia*
- *B. B. Zaidan – PhD Candidate on the Department of Electrical & Computer Engineering / Faculty of Engineering, Multimedia University, Cyberjaya, Malaysia.*
- *Mohamed Elhadi M. Sharif: PhD- Candidate /Universiti Utara Malaysia /MCSE Course.*
- *Hamdan.O.Alanazi – Master Student, Department of Computer System & Technology, University Malaya, Kuala Lumpur, Malaysia*Hide information is the general title for two types of techniques, the first type used to protect information from observers and attackers allegedly writing covered (Covered Writing), or in other words (Steganography), which is the subject of research, and the second type, is used to demonstrate the intellectual property rights (Intellectual Property), or to ensure reliability (Authentication), is called (Digital Watermarking) [9],[10]. The main categories of data hiding are separated into two classes and the classification is shown in Fig 1:

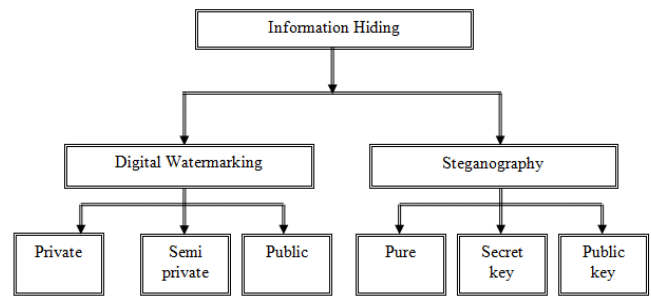

Fig 1.Classfication for Information Hiding

All researchers agree that the term is derived from the Greeks and that there were different points of view on some of the words that came out of this term and include some of them below in relation to the sources [11]. This term derived from Greek, means writing a term covered (Covered Writing), or conceal writing (Hiding Writing). The terminology (Steganography) came from the Greeks and consists of movie (Steganos) means covered or closed and (Graphy) means writing or painting [12]. These mean writing a term covered (Covered Writing). Steganography word can be defined as writing covered (Covered Writing), which is derived from the Greek word [13].

Thus the definition (Steganography) the art of concealment and transfer data through the data again host or Carrier, but harmful harmless transmitters for those data do not allow any enemy or observers to discover that there is confidential data [13].

Digital watermarking provides a represent embedding a message in a piece of digital data without destroying its value. It embeds a known massage in a piece of digital data as a means of identifying the legal owner of the data, these techniques can be used on many types of digital data including still imagery, movies and music [14].

## 1.1 Steganography Protection Against Detection ,Removed

Steganography is derived from the Greek for covered writing and essentially means "to hide in plain sight". As defined by Cachin [1] steganography is the art and science of communicating in such a way that the presence of a message cannot be detected. Simple steganographic techniques have been in use for hundreds of years, but with the increasing use of files in an electronic format new techniques for information hiding have become possible. This document will examine some early examples of steganography and the general principles behind its usage. We will then look at why it has become such an important issue in recent years. There will then be a discussion of some specific techniques for hiding information in a variety of files and the attacks that may be used to bypass steganography. Fig 2 shows how information hiding can be broken down into different areas. Steganography can be used to hide a message intended for later retrieval by a specific individual or group. In this case the aim is to prevent the message being detected by any other party. The other major area of steganography is copyright marking, where the message to be inserted is used to assert copyright over a document. This can be further divided into watermarking and fingerprinting

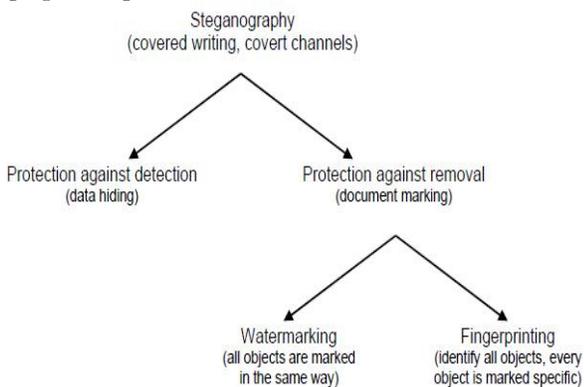

Fig 2. Types of Steganography.

Taken from "An Analysis of Steganographic Techniques" by Popa [2].

Steganography and encryption are both used to ensure data confidentiality. However the main difference between them is that with encryption anybody can see that both parties are communicating in secret. Steganography hides the existence of a secret message and in the best case nobody can see that both parties are communicating in secret. This makes steganography suitable for some tasks for which encryption aren't, such as copyright marking. Adding encrypted copyright information to a file could be easy to remove but embedding it within the contents of the file itself can prevent it being easily identified and removed.

Figure 3 shows a comparison of different techniques for communicating in secret. Encryption allows secure communication requiring a key to read the information. An attacker cannot remove the encryption but it is relatively easy to modify the file, making it unreadable for the intended recipient.
1

|  | Confidentiality | Integrity | Unremovability |
|---|---|---|---|
| Encryption | Yes | No | Yes |
| Digital Signatures | No | Yes | No |
| Steganography | Yes / No | Yes / No | Yes |

Fig 3. Comparison of Secret Communication Techniques.

Taken from "An Analysis of Steganographic Techniques" by Popa [2].

Digital signatures allow authorship of a document to be asserted. The signature can be removed easily but any changes made will invalidate the signature, therefore integrity is maintained.

Steganography provides a means of secret communication which cannot be removed without significantly altering the data in which it is embedded. The embedded data will be confidential unless an attacker can find a way to detect it [3],[4],[5].

## 1.2 Steganography VS Watermarks

Even though both of two technical works using the same principle but there are differ in somehow. Scores of water is to hide the data relatively few "often" what is a legitimate owner of authors digital signature (Digital Authors Signature), documenting the company (Company Logo), the right of reproduction information (Copyright Information) or confidential unique figures (Fingerprinting), and these are all means to establish a reference point downloads the same[12],[13],[14].

The system usually coverage used large amounts of data to be uncovered in other media have nothing to do with the data embedded on the launch, but are a receptacle to contain such data to protect them from discovery [15].

The researchers concern more about Digital watermarking in this project. Thus more literature for next stage.

### 1.3 Digital Watermarking Overview

Digital watermarks are digital data elements that are embedded into actual content—not carried in the header—so the elements survive analog conversion and standard processing, such as conversion to MP3s or changes in file/media format [16]. Digital watermarks may be embedded into, and read from, video, audio and still images to enhance the user experience, facilitate business rules, and enrich the media ecosystem as a whole by allowing all content to be self-identifying or carry information that may trigger a defined behavior. Digital watermarking differs from pattern matching (fingerprinting) in that it is not based on statistical matches against databases of known content. Rather, digital watermarks are the equivalent of placing information within the content itself, enabling detection in stand-alone or connected applications throughout the distribution channel and at play-out [16]. In its most common form, the digital watermark data is not perceptible to the human ear or eye, but can be read by computers. One metric for determining whether a digital watermark is acceptably robust is that, when it is embedded at animperceptible level, it cannot be stripped out without noticeably degrading the host content. The digital data carried by a watermark can consist of any information deemed relevant for a specific application or usage model, but typically falls into two categories: 1) triggers that indicate some action should be taken (e.g., copy control information (CCI), 'flags' or trigger bits); and 2) identifiers that provide information, usually about the content, the distribution service, and/or the player/client (e.g., media serial numbers, service identifier, player or client identifier). Standards for both the structure and the semantics for conveying both categories of information must be agreed upon for a useful P2P watermark detection ecosystem and business regime to develop. Some of this work has already been tackled by other bodies and could be used as a starting point here (e.g., CCI within ATSC and CEA, Media IDs within ISAN) [17],[18].

Digital watermarks are in extensive use around the world, with billions of digitally watermarked objects and hundreds of millions of detectors in use for broadcast monitoring, copy protection, copyright notification, and forensic tracking applications. Major record labels and movie studios currently use digital watermarks to track content in production and prior to release to the public. This effort has led to a significant reduction in illegitimate use of pre-release music and movies, and has resulted in arrests by the FBI of individuals trafficking in screener copies of movies [19].

In addition, a number of digital watermarking providers are helping major content rightsholders in the media and entertainment industry today to mark currently distributed movies and music with media serial numbers. As will be discussed later, this effort is establishing an ecosystem of content that could be leveraged to facilitate the creation of legitimate, new P2P content distribution offerings. Other examples of the use of digital watermarks can be found in the Appendix [19].

## 2. DIFFERENT TYPES OF WATERMARKS

Some of the different types of watermarks that have been developed in the past few years are listed below [16]:

### 2.1 Visible Watermarks

Visible watermarks are designed to be easily perceived by the viewer, and clearly identify the owner; the watermark must not detract from the image content itself, however. Most research currently focuses on invisible watermarks, which are imperceptible under normal viewing conditions [17].

### 2.2 A Watermark May be Fragile, Semi-Fragile or Robust.

A watermark may be fragile, semi-fragile or robust fragile watermarks are designed to be distorted or "broken" under the slightest changes to the image. Semi-fragile watermarks are designed to break under all changes that exceed a user-specified threshold. [A threshold of zero would form a fragile watermark.] Robust watermarks withstand moderate to severe signal processing attacks (compression, rescaling, etc.) on an image [18].

### 2.3 Spatial watermarks

Spatial watermarks are constructed in the image spatial domain, and embedded directly into an image's pixel data. Spectral (or transform-based) watermarks are incorporated into an image's transform coefficients (DCT, Wavelet) [18].

### 2.4 Image-Adaptive Watermarks

Image-adaptive watermarks are usually transform-based, and very robust. They locally adapt the strength of the watermark to the image content through perceptual models for human vision. These models originally developed for image compression [19].

### 2.5 Blind Watermarking Techniques

Blind watermarking techniques can perform verification of the mark without use of the original image. Other techniques rely on the original to detect the watermark.

Many applications require blind schemes; these techniques are often less robust than non-blind algorithms [16].

## 3. WATERMARKING PRINCIPLE

There are some of the steps that are required to embed some digital data in host image to create a watermark image these steps consist of [16]:

- Generating the mark.
- Embedding the mark.
- Creating the key file.
- Producing watermarked image.

They have two main processing encoding and decoding process:

- The encode process is consist of reading the host image (H) and using mark image (W) to generate the watermark image (HW).

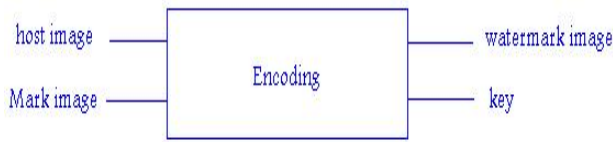

Fig 4. Encoding Watermark Process

- The dencode process is used to extract the mark image from the watermark image by using the key file (Key).

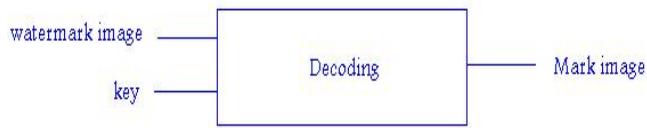

Fig 5. Decoding Watermark Process

## 4. DIGITAL WATERMARK TECHNIQUE REQUIREMENTS

To be really effective for copyright enforcement, a digital watermarking technique must satisfy the following requirements [17]:

### 4.1 Perceptual Transparency

The watermark must be embedded without affecting the perceptual quality of the host media under typical perceptual conditions. That is, human observers cannot distinguish the original host media from the watermarked media. As a result, human eyes should not recognize the existence of the watermark.

### 4.2 Unambiguity

The retrieval of watermark should unambiguously identify the owner. In addition, the accuracy of owner identification should degrade gracefully under attacks.

### 4.3 Robustness

As a watermark is used to identify the owner of digital media, removal of the embedded watermark should be difficult for an attacker or any unauthorized user.

### 4.4 Tamper –Resistance

The embedded watermark must be resistant to tampering through collusion by comparing multiple copies of the media embedded with different watermarks.

## 5. THE PROPOSED ALGORITHM

The embedded watermark must be invisible to human eyes and robust to most image processing operations. To meet these requirements, a bit of binary pixel value (0 or 1) is embedded in block of the host image [18]. Before insertion, the host image is decomposed into NxN blocks.

The sizes of the host and watermark determine the size of the blocks, so in this paper the size of the host image is 512x512 pixel grayscale images with intensity values between 0 and 255 and the watermark is a 128x128 binary image, so the bits for the watermark are embedded into 4x4 blocks (B) of the host image [18].
The algorithm of embedded a signature in the host image need to divide the algorithm in two categories:

### 5.1 Watermark Embedding

After dividing the host image into 4x4 blocks, the steps of insert the bits of the watermark image into each block (B) in the host image are [17]:

- Compute the average, $g_{mean}$, minimum, $g_{min}$, and maximum,

$g_{max}$, of the intensities on the pixels in B.

- Classify each pixel into one of two categories, based on whether its intensity value is above or below the mean of the block. i.e., the $ij^{th}$ pixel, $b_{ij}$ is classified depending on its intensity, g, as

$$b_{ij} \in Z_H \quad \text{if } g > g_{mean}$$
$$b_{ij} \in Z_L \quad \text{if } g \leq g_{mean}$$

Where:
- $Z_H$ and $Z_L$ are the high and low intensity classes, respectively.
1) Compute the means, $m_L$ and $m_H$, for the two classes, $Z_L$ and $Z_H$.

2) Define the contrast value of block B as :

$$C_B = \max(C_{min}, \alpha (g_{max} - g_{min}))$$

Where:
- $\alpha$ is a constant
- $C_{min}$ is a constant which defines the minimal value a pixel's intensity can be modified[19].

- Given the value of the signature image $b_w$ is 0 or 1, modify the pixels in B according to:

  If $b_w = 1$,

  $g_{new} = g_{max}$      if $g > m_H$
  $g_{new} = g_{mean}$      if $m_L \leq g < g_{mean}$

  $g_{new} = g + \delta$      otherwise

  If $b_W = 0$,

  $g_{new} = g_{min}$      if $g < m_L$
  $g_{new} = g_{mean}$      if $g_{mean} \leq g < m_H$

  $g_{new} = g - \delta$      otherwise

Where $g_{new}$ is the new intensity value for the pixel which had original intensity g value and $\delta$ is a random value between 0 and $C_B$.

- The modified block of pixels, $B_{new}$, is then positioned in the watermark image in the same location as the block, B, of pixels from the original host image [20].

These steps describe the procedure by which the watermarked image is generated from a host and a watermark. The pixel intensities are modified within a range specified by the contrast value for a given block. If the contrast value is large, then the pixels are modified more than if the contrast value is small. Thus, the pixels are modified in a manner that is adaptive to the contrast value of the regional block of pixels. The result is that, if a 1 is embedded into a block, the average intensity value for that embedded block will be greater than the average intensity for the same block of the original host image. If a 0 is embedded, then the average intensity of the embedded block will be lower than that of the original host. By using the offset δ, those pixels which it modifies will have a small random noise component, however with a nonzero overall mean value. The random nature of this tuning helps to prevent a visible blocking effect while still contributing to the shift of the overall mean of the block of pixels. This also contributes to the robustness of the algorithm to some of the image filtering processes while also reducing the blocking. The filtering that might be performed on the watermarked image may reduce the variance of the noise; however, given its nonzero mean, the average may still be preserved at a higher level for a given block [20].

### 5.2 Extracting the Embedded Watermark

The extraction algorithm is straightforward and requires retrieving the original host image. The extractor need only compute the sum of the intensity values for the block of the host and watermarked image. A bit is decoded by making the comparison of the two resultant values [17]:

If SW > So' then bW = 1
If SW ≤ So' then bW = 0

Where SW and So, are the sums for the blocks of the watermarked and original images respectively. The decoded bits are then entered into the inverse permuted order as the NxN blocks were selected by using the key from the scrambled insertion procedure. This produces the recovered scrambled watermark. Then, the scrambled watermark is descrambled according to the key from the initial scrambling operation [21].

The decision as to weather there is in fact a watermark in the image is then a subjective one.

### 6. DISCUSSION

Digital watermark refer to the process of embedding in a host information an information mark which is not immediately discernible upon examine the embedded host information these technique have been used as way of reducing counterfeiting in docement . Currency and other application for century's .with the widespread use of digital representation of image, video, audio and other signals, the patent protection or copyright protection by using invisible digital watermarks become a very activity area of research. Naturally many new watermarking applications have become grad interest in this new digital perspective, including national security application such as integrity and authenticity verification, covert communication and traitor tracing finger printing applications. Several other digital watermarking are still emerging, bringing a wide perspective for research.

### ACKNOWLEDGEMENT

Thanks in advance for the entire worker in this project, and the people who support in any way, also I want to thank MMU for the support which came from them.

**Mahmoud Elnajjar:** He obtained his Master degree from Universiti Utara Malaysia at Information technology. His interest area is Datawerhouse and Information Security. Currently, he is PHD- Candidate /Universiti Utara Malaysia / Microsoft Business Intelligence MCSD.NET Developer Solution

**Aos Alaa Zaidan**: He obtained his 1st Class Bachelor degree in Computer Engineering from university of Technology / Baghdad followed by master in data communication and computer network from University of Malaya. He led or member for many funded research projects and He has published more than 50 papers at various international and national conferences and journals, His interest area are Information security (Steganography and Digital watermarking), Network Security (Encryption Methods) , Image Processing (Skin Detector), Pattern Recognition , Machine Learning (Neural Network, Fuzzy Logic and Bayesian) Methods and Text Mining and Video Mining. .Currently, he is PhD Candidate on the Department of Electrical & Computer Engineering / Faculty of Engineering / Multimedia University / Cyberjaya, Malaysia. He is members IAENG, CSTA, WASET, and IACSIT. He is reviewer in the (IJSIS, IJCSNS, IJCSN, IJCSE and IJCIIS).

**Bilal Bahaa Zaidan:** He obtained his bachelor degree in Mathematics and Computer Application from Saddam University/Baghdad followed by master in data communication and computer network from University of Malaya. He led or member for many funded research projects and He has published more than 50 papers at various international and national conferences and journals, His interest area are Information security (Steganography and Digital watermarking), Network Security (Encryption Methods) , Image Processing (Skin Detector), Pattern Recognition , Machine Learning (Neural Network, Fuzzy Logic and Bayesian) Methods and Text Mining and Video Mining. .Currently, he is PhD Candidate on the Department of Electrical & Computer Engineering / Faculty of Engineering / Multimedia University / Cyberjaya, Malaysia. He is members IAENG, CSTA, WASET, and IACSIT. He is reviewer in the (IJSIS, IJCSNS, IJCSN, IJCSE and IJCIIS).

**Mohamed Elhadi M. Sharif:** He obtained his Master degree from Universiti Utara Malaysia at Information technology. His interest area is Datawerhouse and Information Security. Currently, he is PHD- Candidate /Universiti Utara Malaysia /MCSE Course.

**Hamdan Al-Anazi**: has obtained his bachelor dgree from "King Suad University", Riyadh, Saudi Arabia. He worked as a lecturer at Health College in the Ministry of Health in Saudi Arabia, then he worked as a lecturer at King Saud University in the computer department. Currently he is Master candidate at faculty of Computer Science & Information Technology at University of Malaya in Kuala Lumpur, Malaysia. His research interest on Information Security, cryptography, steganography and digital watermarking, He has contributed to many papers some of them still under reviewer.